\def\met{\mbox{${\hbox{$E$\kern-0.6em\lower-.1ex\hbox{/}}}_T$}} 
\def\D0{D\O}                            
\def\d0draft{}
\def\err#1#2#3 {{\it Erratum} {\bf#1},{\ #2} (19#3)}
\def\ib#1#2#3 {{\it ibid.} {\bf#1},{\ #2} (19#3)}
\def\nc#1#2#3 {Nuovo Cim. {\bf#1} ,#2(19#3)}
\def\nim#1#2#3 {Nucl. Instr. Meth. {\bf#1},{\ #2} (19#3)}
\def\np#1#2#3 {Nucl. Phys. {\bf#1},{\ #2} (19#3)}
\def\pl#1#2#3 {Phys. Lett. {\bf#1},{\ #2} (19#3)}
\def\prev#1#2#3 {Phys. Rev. {\bf#1},{\ #2} (19#3)}
\def\prl#1#2#3 {Phys. Rev. Lett. {\bf#1},{\ #2} (19#3)}
\def\rmp#1#2#3 {Rev. Mod. Phys. {\bf#1},{\ #2} (19#3)}
\def\zp#1#2#3 {Zeit. Phys. {\bf#1},{\ #2} (19#3)}

\documentstyle[aps,epsf]{revtex}  

\begin{document}        

\baselineskip 14pt
\title{Recent Electroweak Results from the Tevatron}
\author{Krishnaswamy Gounder}
\address{University of California, Riverside, CA 92521, USA. \\
{\rm (For the CDF and \D0 Collaborations)}}
%
\maketitle              

\begin{abstract}        
     Recent electroweak results from the CDF and \D0 Collaborations at
the Fermilab Tevatron Collider are presented. After a  brief description of
the \D0 measurements of $W/Z$ production cross sections, $W$ width,
$W$ mass and $W \to \tau \nu$ decays, the CDF result on $W(p_T)$
distribution is outlined. The comprehensive search for anomalous gauge
couplings by \D0 in 1992-96 data is presented along with a detailed 
description of the $WW/WZ \to \mu\nu {\rm jj}$ channel.

\end{abstract}   	

\section{Introduction}   

The CDF ~\cite{ref:cdf} and \D0 ~\cite{ref:d0e} collaborations at the 
Fermilab Tevatron collider collected data during Run I (1992-96) 
at $\sqrt{s}=1.8$ TeV 
corresponding to an integrated luminosity of about 130 $pb^{-1}$ for 
each experiment. The large number of $W$ and $Z$ bosons detected in
the electron and muon channels were used to make precise measurements
of their properties. The $W \to \tau\nu$ decay was studied by both 
experiment to measure the ratio $g^W_{\tau}/g^W_e$. Using the Run I
data, \D0 has made a comprehensive search for anomalous trilinear gauge
couplings in 12 different diboson channels and has combined
them to produce some of the most stringent anomalous gauge coupling 
limits in the world so far.

Run I was divided into three different sections: Run 1A -
1992-93 ($\sim\ 20\ pb^{-1}$); Run 1B - 1993-95 ($\sim\ 90\ pb^{-1}$);
Run 1C - 1995-96 ($\sim\ 20\ pb^{-1}$). After a brief review of $W$
and $Z$ properties in Section II, the \D0 search for anomalous gauge 
couplings in the diboson final states is presented in Section III.

\section{Electroweak Measurements}

\subsection{$W$ and $Z$ Boson Production }

Due to cleaner signatures and lower backgrounds, the $W$ and $Z$ bosons
are detected via their leptonic decays: $W \to e\nu, \mu\nu$ and $Z
\to ee, \mu\mu$. The $W$ event selection requires an electron (muon)
with $p_T >$ 25 (20) GeV/c and {\met} $>$ 25 (20) GeV in the event. 
For $Z$ selection, the {\met} requirement is replaced by that of a 
similar second lepton. The backgrounds for the $W$ electron sample are
mainly due to QCD fakes (5.7\%), $\tau$ decays (1.8\%) and one-legged 
$Z$ decays (0.6\%). The total background for the $W$ $\mu$ sample
is 19.8\% including cosmic muons. The $e$ and $\mu$ $Z$ samples 
contain 4.8\% and 11.6\% backgrounds respectively with additional 
contributions from Drell-Yan pairs and combinatorics.

\begin{figure}
\centerline{
\begin{tabular}{c c}
\epsfxsize 3.0 truein
\epsfysize 2.0 truein
{\epsfbox{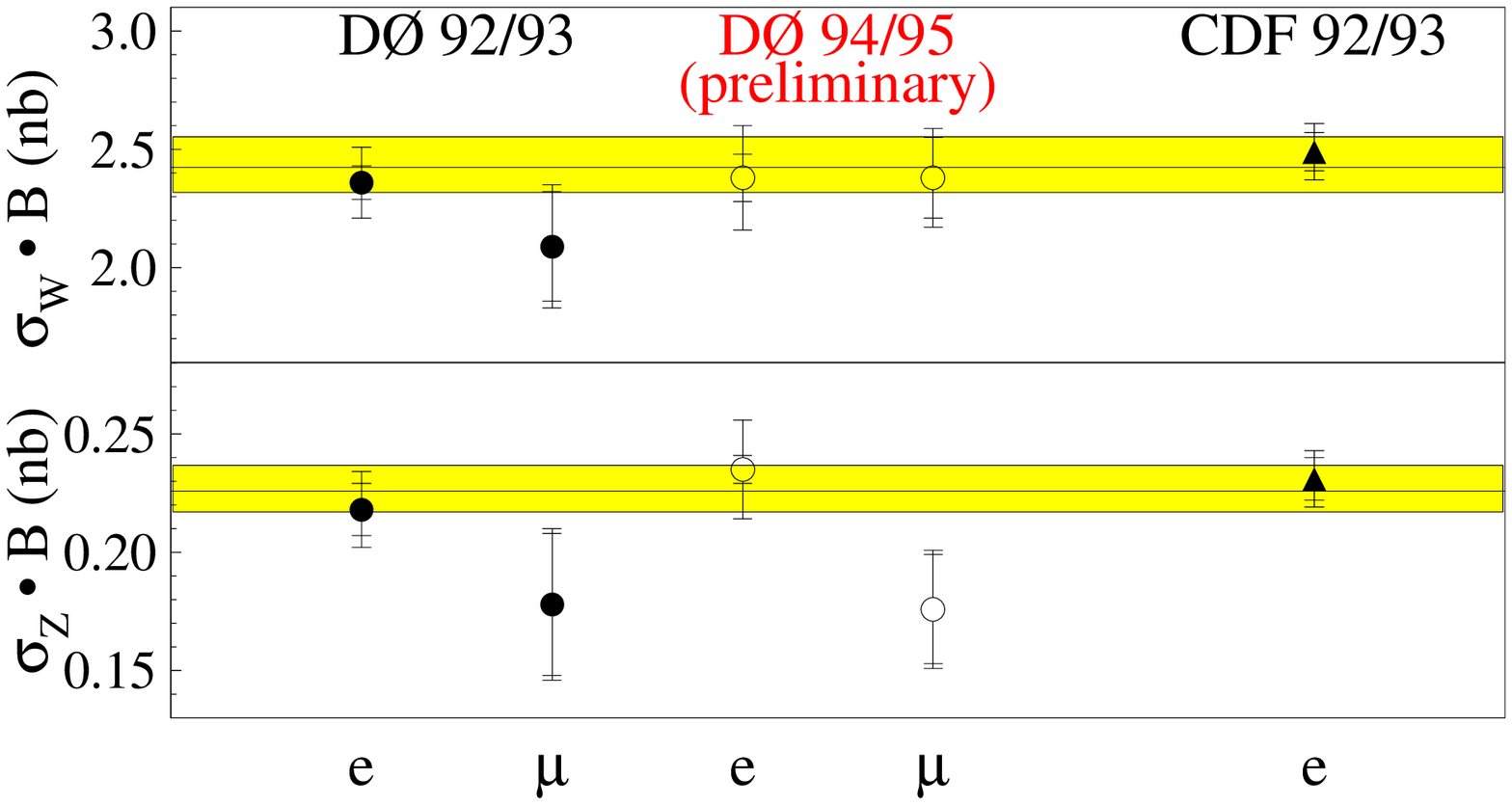}} &
\hspace*{0.6in}
\epsfxsize 2.5 truein
{\epsfbox{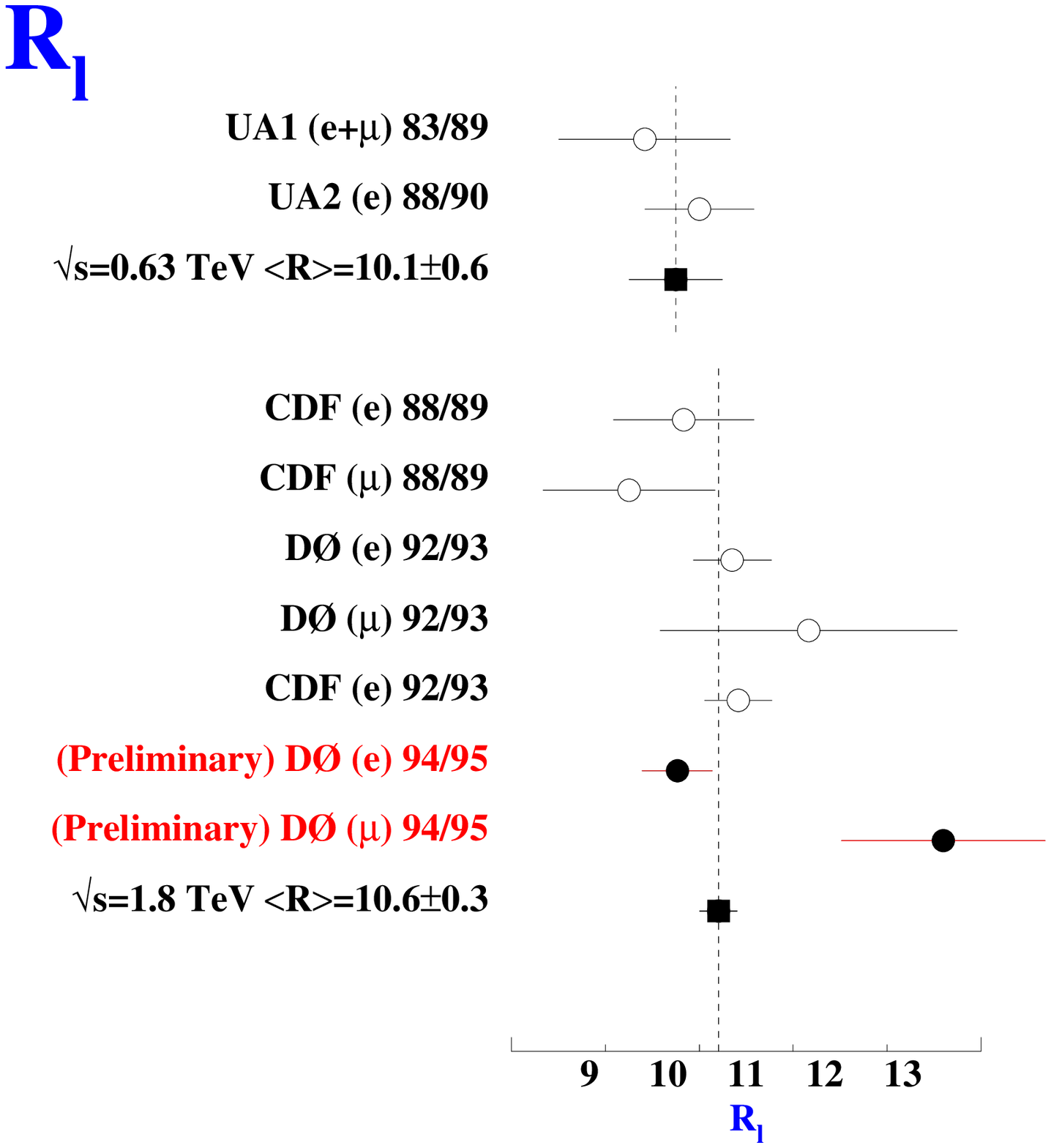}}
\end{tabular}
}
\caption{The Tevatron measurements of $\sigma \cdot B$ for inclusive $W$
and $Z$ production are shown on the left. The shaded bands are
the ${\displaystyle O}(\alpha_s^2)$ theoretical QCD predictions. On
the right, a summary of $R_l$ measurements including the Tevatron 
average.}
\label{fig:wzprod}
\end{figure}

The recent $W$ and $Z$ cross section results from CDF and \D0 are 
shown in Figure~\ref{fig:wzprod} and are compared to the 
${\displaystyle O}(\alpha_s^2)$
theoretical QCD prediction ~\cite{ref:hamvan}. When the ratio of $W$
and $Z$ cross sections is computed many uncertainties cancel and an 
indirect measurement of the $W$ width can be made: 
$$R_l = \frac{\sigma \cdot B(W \to l\nu)}{\sigma \cdot B(Z \to ll)} = 
\frac{\sigma(W)}{\sigma(Z)}\frac{\Gamma(W \to l\nu)}{\Gamma(W)}
\frac{1}{BR(Z \to ll)}$$
With the measured ratio of $R_l = 10.6\pm0.3$, a value of 
$\Gamma(W) =2.06\pm0.06$ is obtained from the above equation by 
using the a theoretical calculation of $\sigma_W/\sigma_Z$,
the precise measurement of $B(Z \to ll)$ from LEP, and the theoretical
computation of $\Gamma(W \to l\nu)$. A comparison of $R_l$ 
measurements is shown in Figure~\ref{fig:wzprod} along with the
Tevatron average. An updated and expanded $W/Z$ production results 
can be found in reference~\cite{ref:seno}.

\subsection{$W$ Mass }

The $W$ boson mass is a fundamental parameter of the Standard
Model. In the on shell scheme:
$$M_W = \left(\frac{\pi\alpha(M^2_Z)}{\sqrt{2}G_F}\right)^{1/2}
\frac{1}{\sin\theta_W \sqrt{1-\Delta r}}$$ 
where $M_Z$ is the Z boson mass, $\alpha$ is the fine structure
constant, $\theta_W$ is the weak mixing angle, $G_F$ is the Fermi coupling
constant and $\Delta r$ denotes the radiative corrections. The $\Delta
r$ is sensitive to masses of particles such as the Higgs boson, top
quark and other new particles. Therefore, a precision measurement of
the $M_W$ can be used for constraining the Higgs mass or probing the 
presence of new physics beyond the SM ~\cite{ref:wnew}.

\vskip -0.3in

The $W$ mass is measured using the decay $W \to e\nu$ with a total
luminosity of $\int {\cal{L}}dt= 80 pb^{-1}$. As an 
unknown amount of energy goes down the beam pipe in the forward
and backward directions, the $p_L(\nu)$ remains uncertain. Hence,
the $W$ mass is determined using a likelihood fit to the transverse 
mass $M_T(e\nu) = {[2p^e_Tp^\nu_T(1-cos\phi^{e\nu)}]^{1/2}}$. A similar
procedure is applied to $p_T(e)$ and $p_T(\nu)$ as cross checks.

\begin{figure}
\epsfxsize 2.5in
\epsfysize 2.5in
{\epsfbox{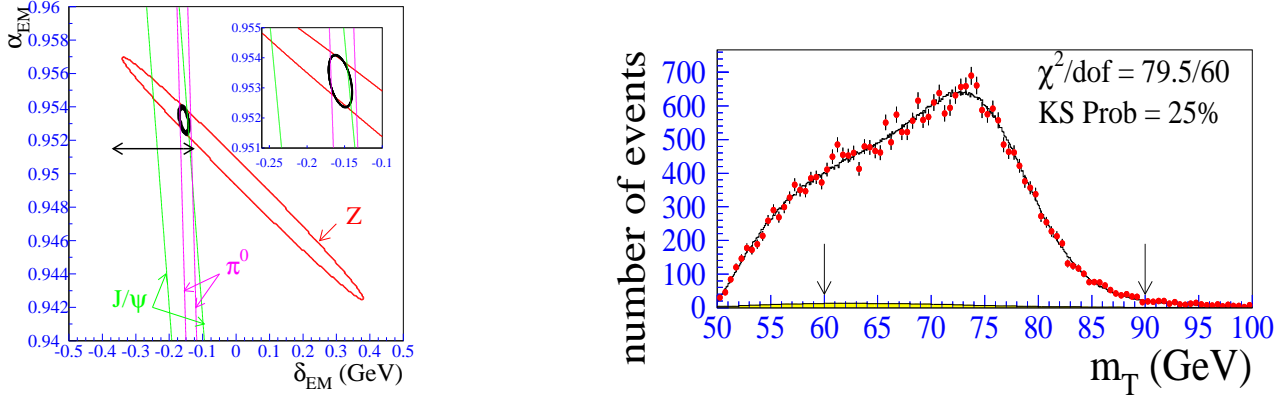}}
\caption{The left figure shows the constraints on the \D0
electromagnetic energy scale parameters from collider data. The
transverse mass distribution of $W \to e\nu$ events from \D0 Run IB
data with the best fit superimposed is shown on the right.}
\label{fig:wmass1}
\end{figure}

The event selection and backgrounds are similar to the cross section
measurement. Event selection with a high quality isolated electron 
in the central region with $p_T(e) > 25$ GeV/c, {\met} $>$ 25 GeV
and hadronic recoil $<$ 15-20 GeV/c leads to a final sample of about
28,000 events. The $p_T(\nu)$
depends on the recoil momentum of the electron and hadrons which
relies on the detailed understanding and modelling of the 
leptonic and hadronic energy scales. The \D0 electromagnetic energy
scale was calibrated using the constraints from decays $Z \to ee,\ 
J/\psi \to ee$ and $\pi^0 \to \gamma\gamma$ as shown in
Figure~\ref{fig:wmass1} where $\alpha$ and $\delta$ are given by 
$E_{meas} = \alpha E_{true} + \delta$. A complete list of uncertainities
contributing to the $W$ mass measurement is listed in
Table~\ref{tab:wmass}.

\begin{figure}
\centerline{
\begin{tabular}{c c}
\epsfxsize 2.5 truein
{\epsfbox{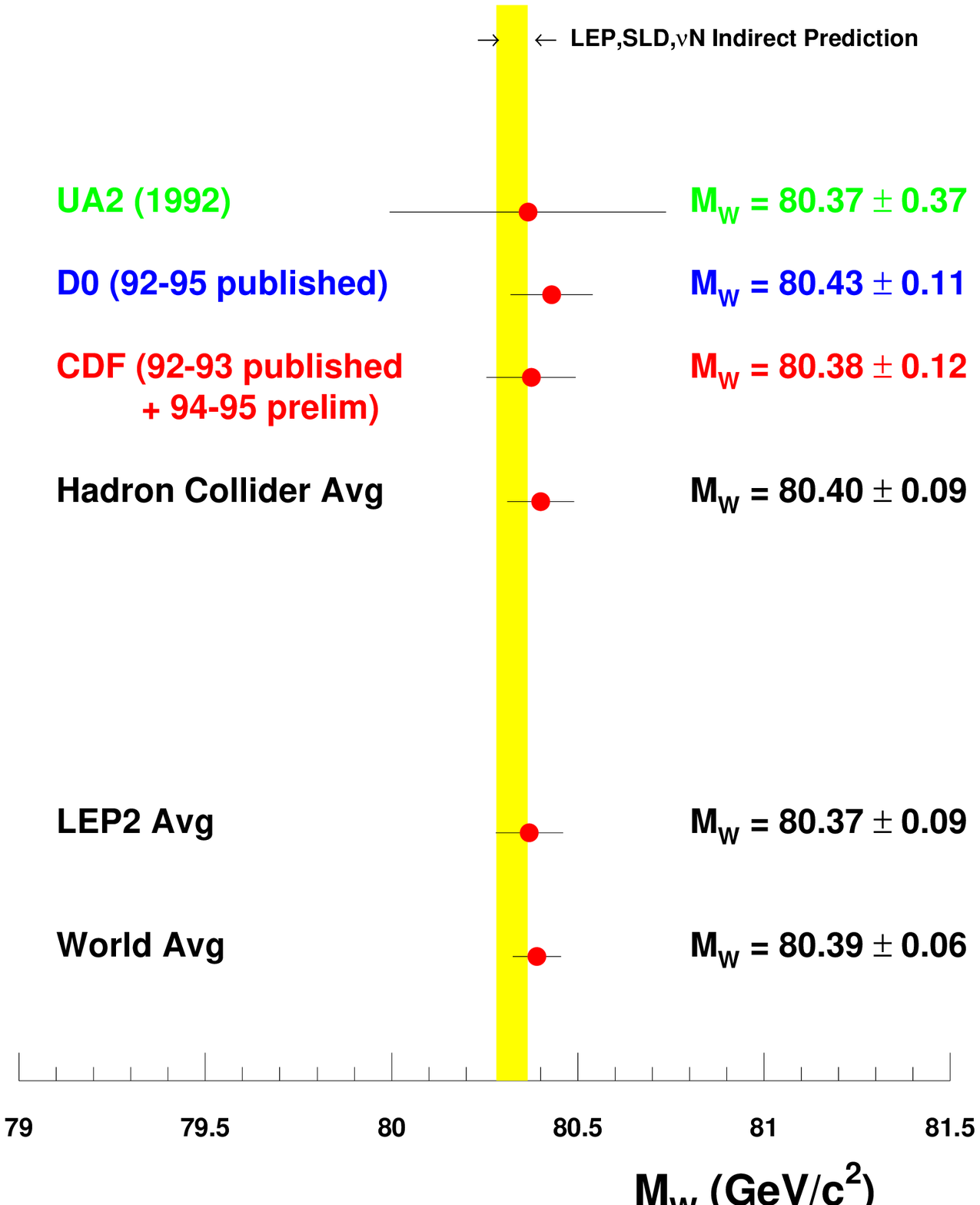}} &
\hspace*{1.0in}
\vspace*{0.5in}
\epsfxsize 2.7 truein
{\epsfbox{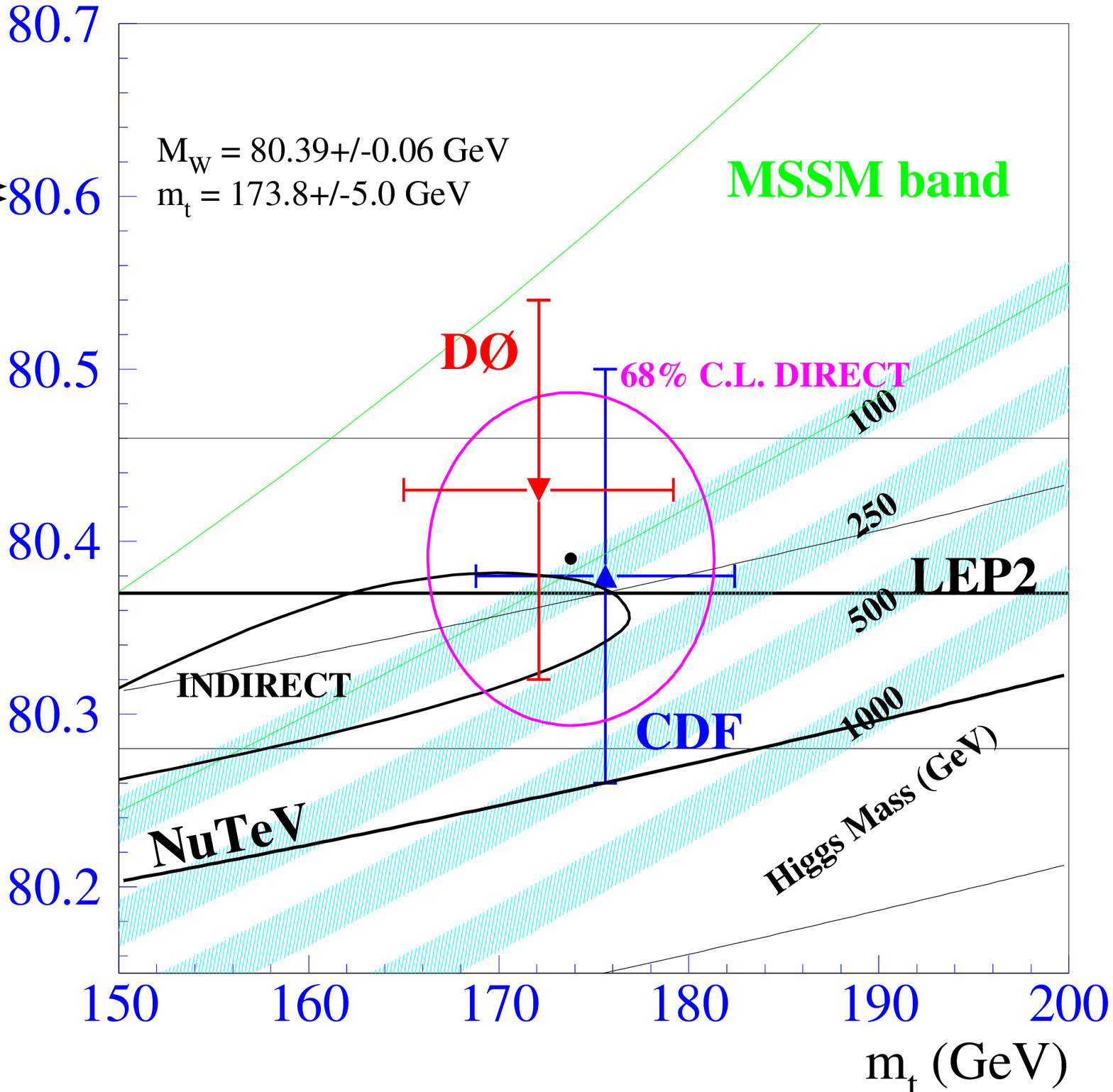}}
\end{tabular}
}
\vspace* {-0.2in}
\caption{A summary of direct $M_W$ measurements is shown on the left
including the world average. The right figure illustrates the
constraints on the Higgs mass imposed by the present Tevatron $M_W$
and $M_t$ measurements as well as by the ones from LEP, NuTeV
and SLC. The bands represent the prediction of the MSSM.}
\label{fig:wmass2}
\end{figure}

The fit to the $M_T(e\nu)$ distribution is shown in
Figure~\ref{fig:wmass1}. The arrows indicate the fitted region for the
final measurement and the shaded region represents background in the 
sample. The results from the $M_T(e\nu)$ and $p_T(e)$
fits using the Run 1B data are $80.44\pm0.01\pm0.07$ and
$80.48\pm0.11\pm0.14$ GeV/$c^2$ respectively. Combining with the 1A
measurement, \D0 1A + 1B: $M_W = 80.43\pm0.11$ GeV/$c^2$. A summary of
direct $M_W$ measurements is shown in Figure~\ref{fig:wmass2} along
with the world average. Figure~\ref{fig:wmass2} also illustrates
the constraints imposed by the direct measurements of $M_W$ and $M_t$
by  CDF and \D0 on the mass of the Higgs boson. Also shown are the 
indirect SLC/LEP2 and NuTev~\cite{ref:nutev} measurements and the 
prediction of the Minimal Supersymmetric Model (MSSM). An update
\D0 $W$ mass measurements including forward detectors can be found in
reference~\cite{ref:seno}.

\begin{table}
\caption{Summary of uncertainities and their sources in the measurement
of $M_W$ using a maximum likelihood fit of $M_T(e\nu)$ and $p_T(e)$
distributions (in MeV/$c^2$).}
\label{tab:wmass}
\begin{tabular}{lcclcc}
{Source} & {$M_T$ Fit} & {$p_T(e)$ Fit} & 
{Source} & {$M_T$ Fit} & {$p_T(e)$ Fit} \\ \tableline \tableline
{$W$ Statistics} & {70} & {85} &
{Electron Energy Scale} & {65} & {65}\\ \tableline
{Calorimeter Linearity} & {20} & {20} &
{Calorimeter Uniformity} & {10} & {10} \\
{Electron Energy Resolution} & {25} & {15} &
{Electron Angle Calibration} & {30} & {30} \\
{Electron Removal} & {15} & {15} &
{Selection Bias} & {5} & {10} \\
{Hadronic Recoil Modelling} & {30} & {20} &
{Input $p_T(W)$ and PDF's} & {25} & {70} \\
{Radiative Decays} & {15} & {15} &
{Backgrounds} & {10} & {20} \\ \tableline
{Total Statistical} & {95} & {105} & 
{Total Systematics} & {70} & {90} \\ 
\end{tabular}
\end{table}

\vspace* {-3.2in}
\begin{figure}
\hspace* {3.0in}
\epsfxsize 4.0in
\epsfysize 4.5in
{\epsfbox{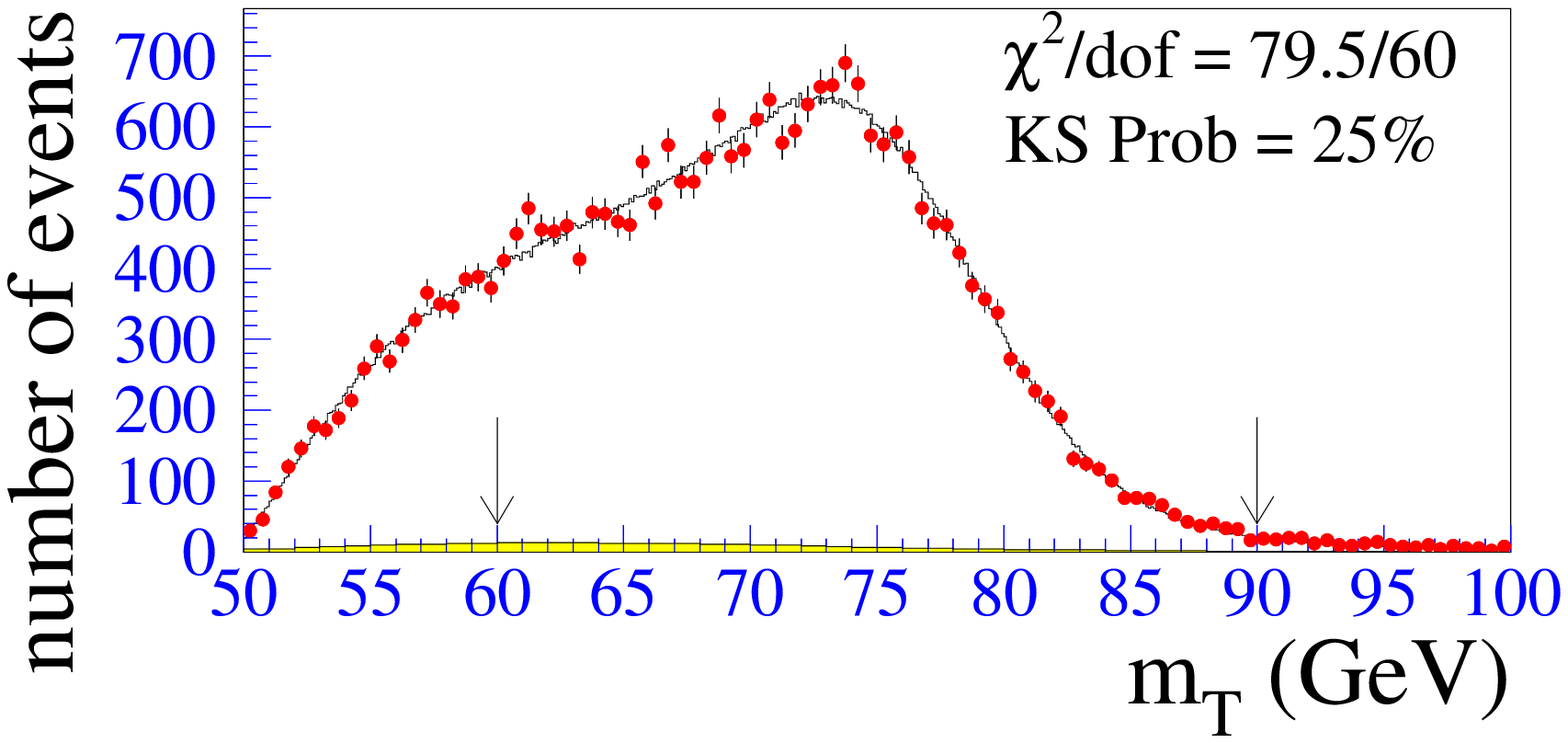}}
\end{figure}
\vskip -2,0in

\subsection{$W \to \tau\nu$ and $g^W_\tau /g^W_e$ }

Both CDF and \D0 have measured $W$ production in the mode  
$W \to \tau\nu$ where $\tau$ decays hadronically. The signature of  
hadronic decay of the $\tau$ is an isolated narrow jet composed
of highly boosted decay products. The event selection requires
$E_T(jet) > 25$ GeV/c and {\met} $>$ 25 GeV. The signal and QCD
backgrounds ($<$ 20\%) are estimated using the $E_T(jets)$ profile
distribution (fractional {\met} in the two hottest towers of the
$\tau$ jet). The \D0 preliminary results are 
$\sigma \cdot BR(W \to \tau) = 2.38 \pm 0.09\ (stat) \pm 0.10\ (sys) \pm
0.1\ (lum)$ nb and $g^W_\tau/g^W_e = 1.004 \pm 0.032$.

The CDF collaboration having a silicon vertex detector has used the 
decay $\tau \to e\nu\nu$ to measure $R_{BR} = BR(W \to
\tau\nu)/BR(W \to e\nu)$. The technique is based on the difference in 
electron impact
parameter distributions in a sample of single electron events composed
of $W \to e\nu$ decays, $W \to \tau\nu,\ \tau \to e\nu\nu$ decays and
QCD backgrounds. A likelihood fit to the branching fraction $R_{BR}$
is performed using $f(d_0;b,c) = a f_e(d_0) + b f_\tau(d_0) + c
f_{bkg}(d_0)$ where the relevant impact parameter distributions
$f_e(d_0),\ f_\tau(d_0)$ are extracted from Monte Carlo and
$f_{bkg}(d_0)$ from QCD data. The results are $R_{BR} = 1.03
^{+0.38}_{-0.32} \pm 0.18$ and $\ g^W_{\tau}/g^W_e = 1.01 \pm 0.17 \pm
0.09$ consistent with hadronic $\tau$ decay measurements and $\tau-e$
universality.

\subsection{$W(p_T)$ Distribution }

The CDF collaboration has measured the $p_T(W)$ distribution in the
decay mode $W \to e\nu$ using Run I data (${\int}{\cal{L}}dt 
= 110\ pb^{-1}$). The measurement goals are to test
perturbative QCD at large $p_T$ ($>$ 20 GeV/c) and differentiate gluon 
resummation techniques ~\cite{ref:resum} (such as q-t-space {\sl vs} b-space)
at small $p_T$ ($<$ 10 GeV/c). Moreover the knowledge of the
$W/Z$ production is essential for the search for new physics and 
is a source of systematics for the measurements of $M_W,\ \Gamma_W$, 
etc. The event selection demands an electron with $p_T(e)\ge$ 25 GeV/c 
and {\met} $\ge$ 25 GeV. The major backgrounds ($<$ 15\%) are due to QCD
electron fakes, $W \to \tau\nu \to e\nu\nu\nu$ and the one-legged
decays of $Z \to ee$.
\vskip -0.9in
\begin{figure}
\epsfxsize 3.8in
\epsfysize 5.3in
\vspace* {-0.25in}
{\epsfbox{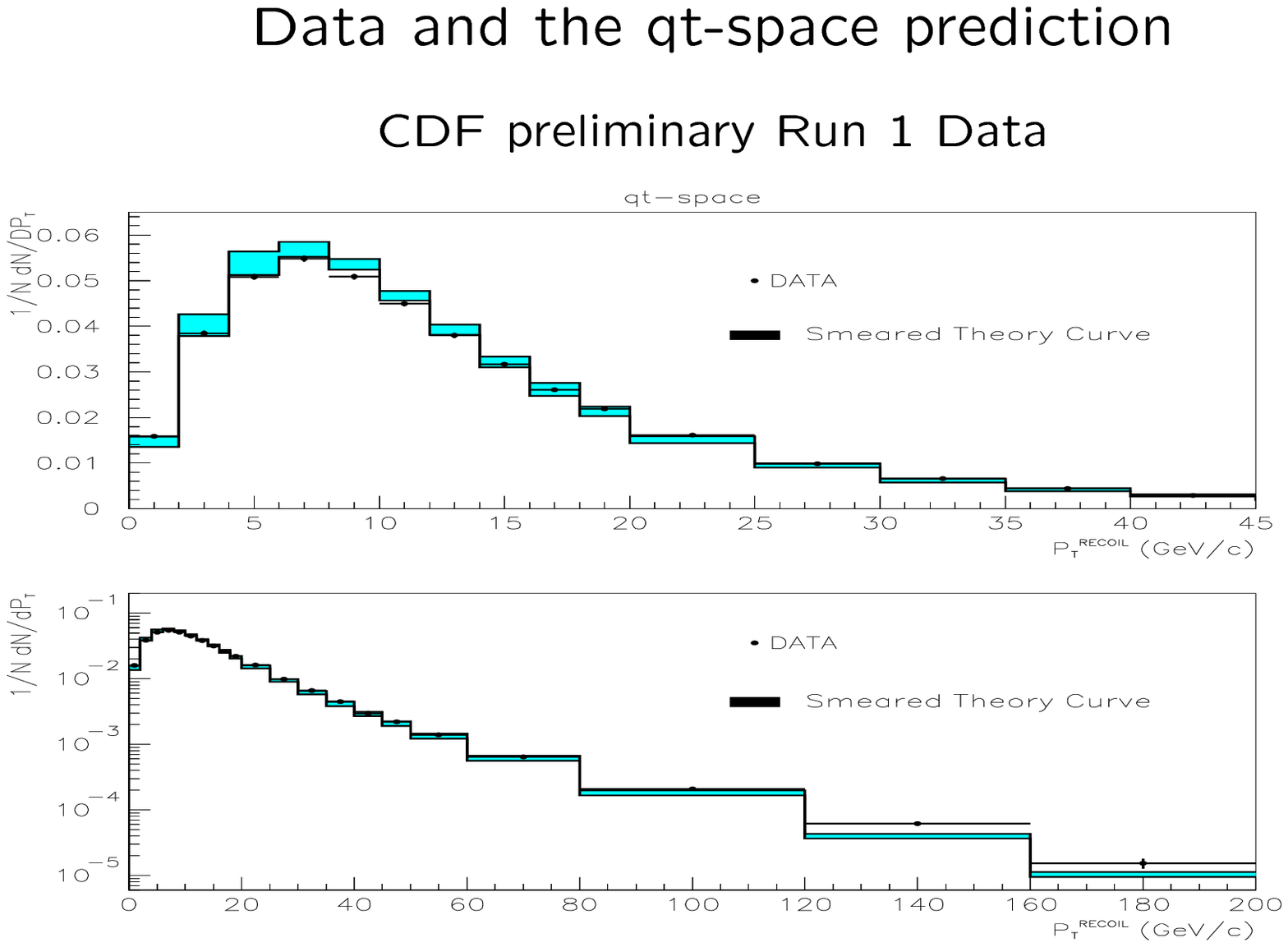}}
\end{figure}
\vspace* {-4.95in}
\begin{figure}
\hspace* {3.7in}
\epsfxsize 3.2in
\epsfysize 3.2in
{\epsfbox{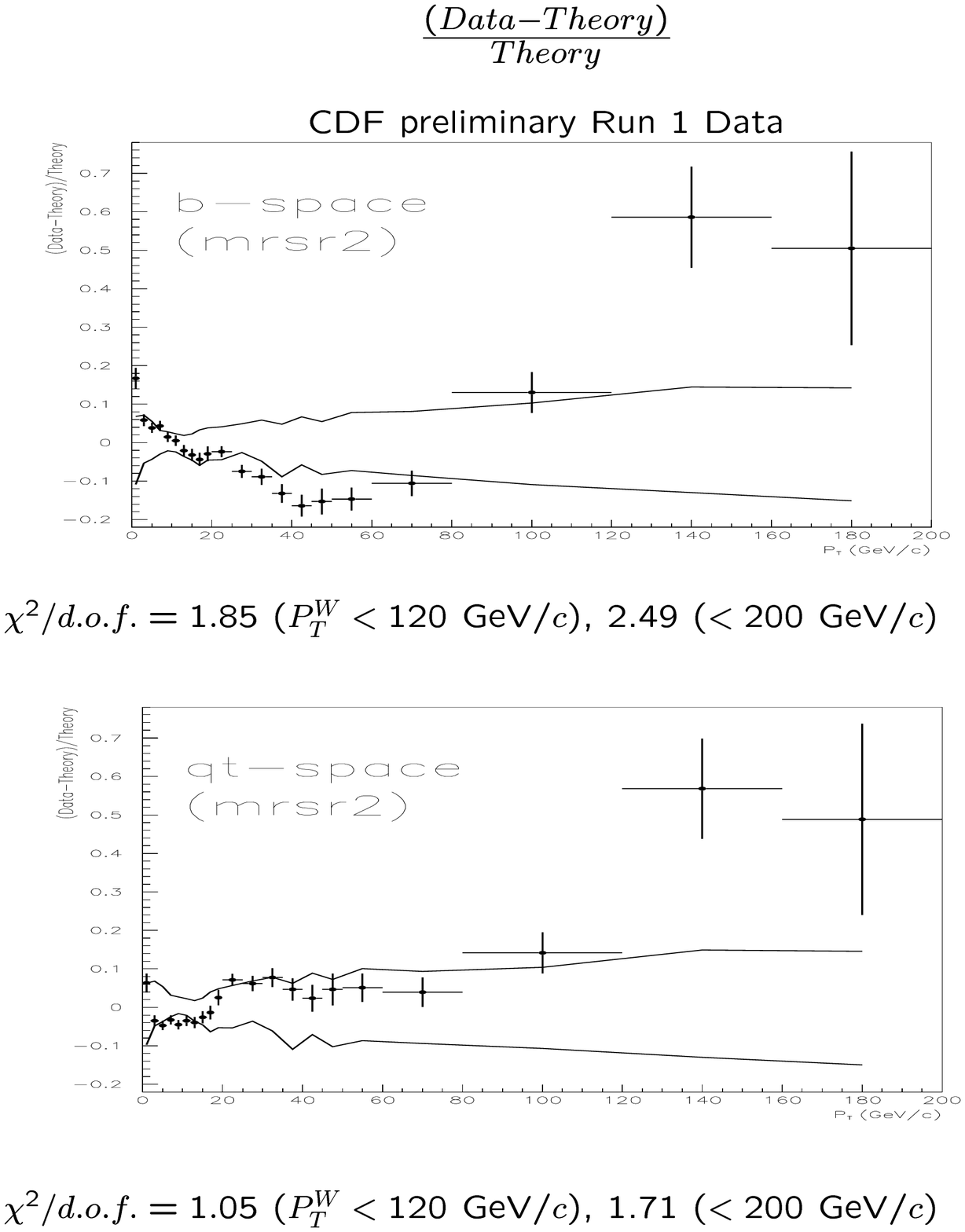}}
\caption{The comparison of data and theoretical predictions using the
qt-space gluon resummation technique for $p_T(W)$ distribution (CDF)
is shown on the left. On the right, the ratio (Data-Theory)/Theory is
plotted as a function of $p_T(W)$ using the q-t-space and b-space gluon
resummation methods.}
\label{fig:cdfpt}
\end{figure}

The $p_T(W)$ distribution using qt-space gluon resummation technique
is shown in Figure~\ref{fig:cdfpt}. Figure~\ref{fig:cdfpt} also displays 
the ratio (data-theory)/theory {\sl vs} $p_T(W)$ for both resummation 
techniques. 
The theory distributions used are obtained by adding the single boson
backgrounds and are smeared with detector resolutions while the
data distributions are after the subtraction of the QCD background. 
Both resummation techniques agree with data up to $p_T(W) \approx$ 120
GeV. At higher $p_T(W)$ region, there appears to be some discrepancy
as in the case of CDF $Z \to ee$ data ~\cite{ref:pvt}. The recently
completed \D0 $W(p_T)$ measurements can be found in 
reference~\cite{ref:wptd0}.

\section{Search for Anomalous Gauge Couplings}

\begin{figure}
\hspace* {3.7in}
\epsfxsize 3.0in
\epsfysize 2.6in
{\epsfbox{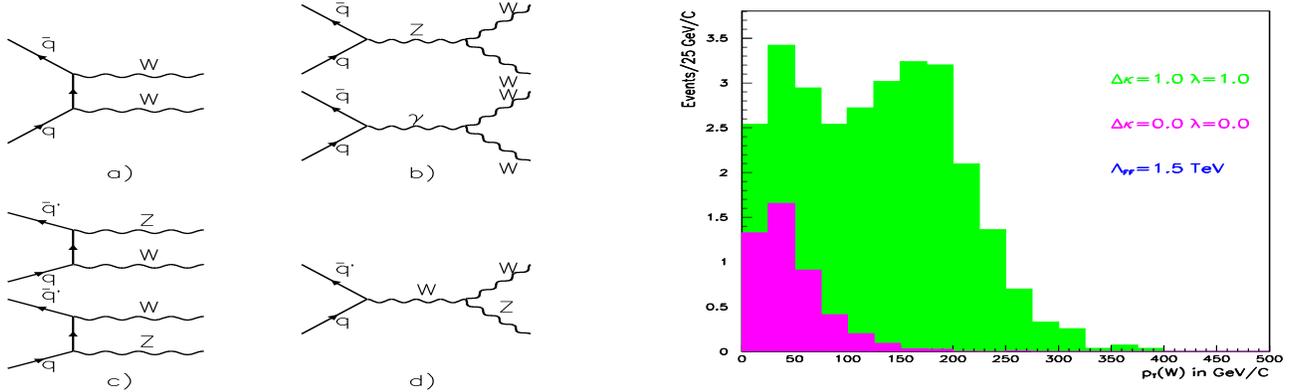}}
\vspace* {-0.2in}
\caption{The tree level Feynman diagrams for $p{\overline p} \to
WW/WZ$ production at the Tevatron are shown on the left. The right
figure shows the comparison of the Standard Model and Anomalous Gauge
Coupling predictions for ${\int}{\cal{L}}dt{\ }= {\ }80.7{\ }pb^{-1}$
at the \D0 Run I detector in the channel $WW/WZ \to \mu\nu {\rm jj}$.}
\label{fig:agc1}
\end{figure}

The effective CP conserving Lagrangian for the description of
anomalous gauge couplings (AGC) is given by ~\cite{ref:agcl}:
$iL_{eff}^{WWV} =
g_{WWV}{\{} g_1^V(W_{\mu\nu}^\dagger W^\mu-W^{\dagger\mu}W_{\mu\nu})
V^\nu + \kappa_V W_\nu^\dagger W_\nu V^{\mu\nu}
 + (\lambda/m_W^2)W_{\rho\mu}^\dagger W_\nu^\mu 
V^{\nu\rho} \\
{\hspace*{3.8in}}+ ig_5^V \epsilon_{\mu\nu\rho\sigma}[(\partial^\rho 
W^{\dagger\mu}) W^\nu - W^{\dagger\mu} (\partial^\rho
W^\nu)]V^\sigma{\}}$ \\
where V = Z, $\gamma$. If CP violating terms are allowed,
three additional terms will appear in the above Langrangian. The 
overall couplings are defined as $g_{WW\gamma}$ = e
and $g_{WWZ}$ = e cot$\theta_W$. At tree level in the SM, the parameters 
are uniquely determined: $g_1^Z = g_1^\gamma = \kappa_Z = \kappa_\gamma
= 1,  \lambda_Z = \lambda_\gamma = g_5^Z = g_5^\gamma = 0$. For
on-shell photons, $g_1^\gamma = 1$ and $g_5^\gamma = 0$ are fixed by
electromagnetic gauge invariance while $g_1^Z$ and $g_5^Z$ may, however,
differ from their SM values. The deviations from tree level SM values
can be cast as: $\Delta g_1^Z = (g_1^Z-1), \Delta\kappa_\gamma =
(\kappa_\gamma-1), \Delta\kappa_Z = (\kappa_Z-1), \lambda_\gamma,
\lambda_Z, g_5^Z$. Most theoretical arguments suggest that these
anomalous couplings are significant at $O(m_W^2/\Lambda^2)$ where
$\Lambda$ is the scale of new physics. To avoid unitarity violations
the coupling parameters should be expressed as form factors such as: 
$\lambda(\hat{s}) = {\lambda}/{(1 + \hat{s}/\Lambda_{FF}^2)^2};\ 
\Delta\kappa(\hat{s}) = {\Delta\kappa}/{(1 +
\hat{s}/\Lambda_{FF}^2)^2}$ with $\Lambda_{FF}$ as the form factor
scale. The tree level Feynman diagrams for
$p{\overline p} \to WW/WZ$ production at the Tevatron are shown in 
Figure~\ref{fig:agc1}.

     Limits on these couplings are usually obtained under the
assumption of equal couplings for $WW\gamma$ and $WWZ$ $(g_1^\gamma =
g_1^Z = 1, \Delta\kappa_\gamma = \Delta\kappa_Z$, and $\lambda_\gamma
= \lambda_Z)$. In the literature, another set of relations is
frequently used, the HISZ relations ~\cite{ref:hisz} where the $WWZ$ 
and $WW\gamma$ couplings are related by: $\Delta\kappa_Z = \Delta\kappa_\gamma 
(1-\tan^2\theta_W)/2;\ \Delta g_{1Z} = \Delta\kappa_\gamma/2\cos^2\theta_W; 
\ \lambda_Z = \lambda_\gamma$. 

     The evidence for non SM physics contributions will be an
enhancement in the high $p_T(W)$ region ~\cite{ref:kggg}. 
Figure~\ref{fig:agc1} illustrates the predicted number of events 
in the mode $WW/WZ \to \mu\nu {\rm jj}$ (for Run 1B) {\sl vs} the $p_T(W)$ 
as a comparison of SM and non SM physics (for $\lambda = 1$ and 
$\Delta\kappa = 1$). Therefore, a study of the $p_T(W)$
spectrum of $WW/WZ$ production will provide a sensitive test of the $WWZ$
and $WW\gamma$ couplings.

\vskip -0.1in
\begin{figure}
\epsfxsize 3.0in
\epsfysize 2.2in
{\epsfbox{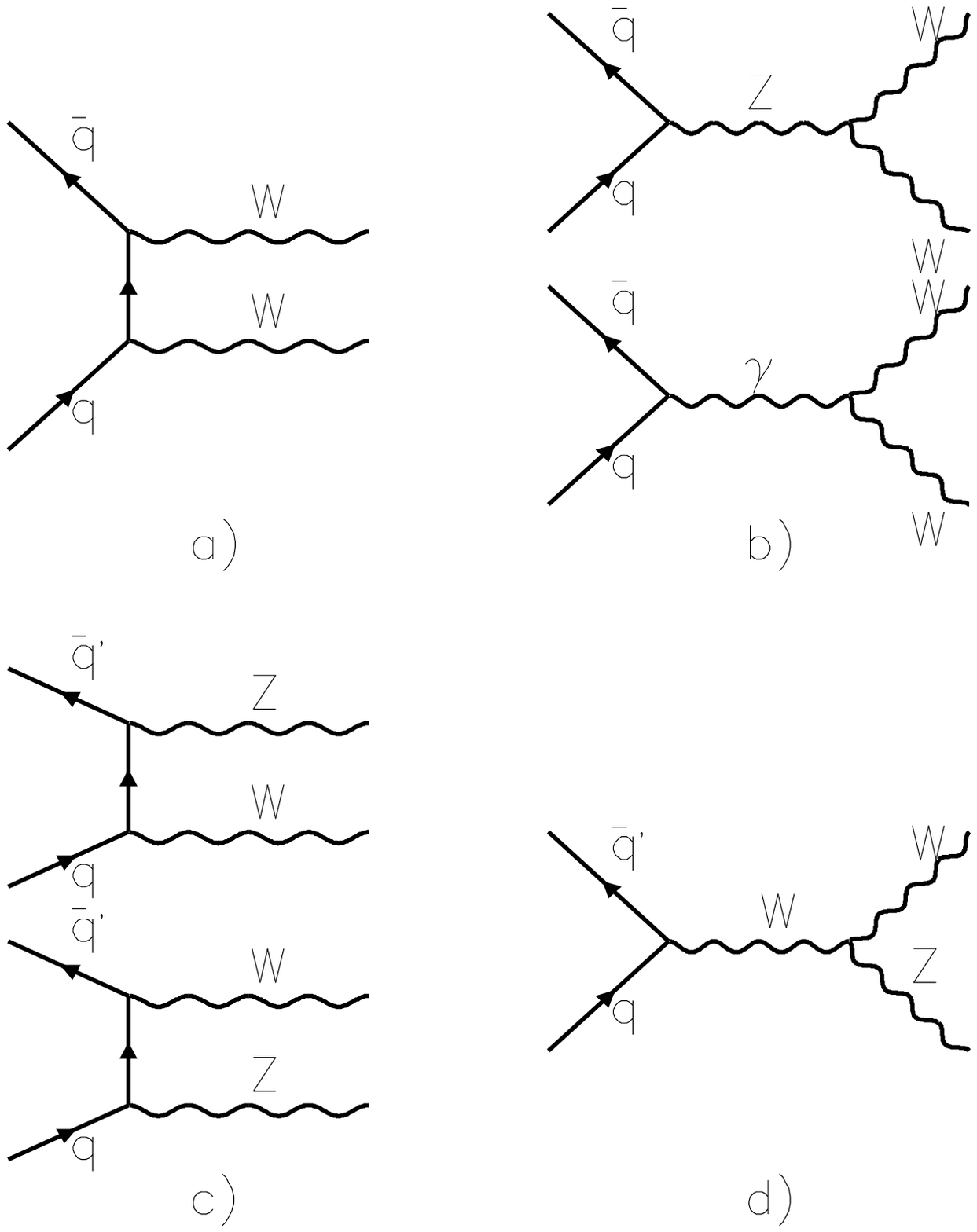}}
\end{figure}
\vspace* {-2.7in}

\subsection{ $WW/WZ \to \mu\nu{\rm jj}$}

The 1994-95 data with ${\int}{\cal{L}}dt{\ }= {\ }80.7{\ }pb^{-1}$ were
used in the search for anomalous $WW/WZ$ production in the mode $WW/WZ
\to \mu\nu {\rm jj}$. The event selection requires a high $p_T$ muon
($>$ 20 GeV/c), {\met} $>$ 20 GeV and at least two good jets with 
$p_T(jet) >$ 20 GeV/c. There is no differentiation between the two
processes $W \to {\rm jj}$ and $Z \to {\rm jj}$ due to limited mass
resolution of the calorimeter. The invariant mass of the two highest
$E_T$ jets in the event is required to be between 50 and 110 GeV/$c^2$
along with a constraint on the transverse mass 
$M_T(\mu\nu)= [2\cdot E^\mu_T\cdot\met\cdot (1- 
\cos(\phi_\mu-\phi_\nu))]^{1/2}>$ 40
GeV/$c^2$. The final event count is $224\pm15$ while the SM prediction
is $4.5\pm0.8$ events. 

The major sources of backgrounds are from QCD multijet and $W
+ \ge 2$ jet events with $W \to \mu\nu$. The QCD multijet background
is due to misidentifying a muon contained in one of the jets as an 
isolated muon and where there is significant {\met}. This background 
is estimated from data using a control sample to determine muon fake
probability~\cite{ref:fake}. The $W + \ge 2$ jets 
contribution is computed
using a Monte Carlo sample generated with VECBOS,
HERWIG~\cite{ref:herw} 
(hadronization) and 0GEANT~\cite{ref:gean} for 
detector simulation. Normalization of 
this background is determined by comparing the number of expected
events outside the dijet mass window after the subtraction of the QCD
multijet contribution. The final background contributions are
$105\pm19$ (QCD multijet), $117\pm24$ ($W + \ge 2$ jets) and
$2.7\pm1.2$ (others) without systematics. The total background
contribution to the final sample are $224.5\pm32.7\ \pm45.8$. 

\begin{table}
\caption{The \D0 Run IB $WW/WZ \to \mu\nu {\rm jj}$ axis limits
(one-dimentional) at the 95\% confidence limits (C.L.) with the assumption of
equal $WW\gamma$ and $WWZ$ couplings and the HISZ relations. The limits
are listed for two different
values of $\Lambda_{FF}$ along with the relevant unitary bounds.}
\label{tab:mujet}
\begin{tabular}{lcccc}
{95{\%} C.L. Limits} & $\Lambda_{FF}=1.5$ TeV & {Unitary Bounds} & 
{$\Lambda_{FF} = 2.0$ TeV } & {Unitary Bounds} \\ \tableline
$\lambda_\gamma=\lambda_Z{\ }(\Delta\kappa_\gamma=\Delta\kappa_Z
= 0$) & { -0.45, 0.46} & {-0.82 0.82} & {-0.43, 0.44} & {-0.46, 0.46} \\ 
{$\Delta\kappa_\gamma=\Delta\kappa_Z{\ }(\lambda_\gamma=\lambda_Z
= 0$)} & { -0.62, 0.78} & {-1.17, 1.17} & {-0.60, 0.74} & {-0.66,
0.66} \\ \tableline
{$\lambda_\gamma({\rm HISZ}){\ }(\Delta\kappa_\gamma
= 0$)} & { -0.44, 0.46} & {-0.82, 0.82} & {-0.42, 0.44} & {-0.46,
0.46} \\ 
{$\Delta\kappa_\gamma({\rm HISZ}){\ }(\lambda_\gamma
= 0$)} & {-0.75, 0.99} & {-2.17, 2.17} & {-0.71, 0.96} & {-1.22, 1.22}
\\
\end{tabular}
\end{table}

\begin{figure}
\hspace* {3.8in}
\epsfxsize 2.5in
\epsfysize 2.5in
{\epsfbox{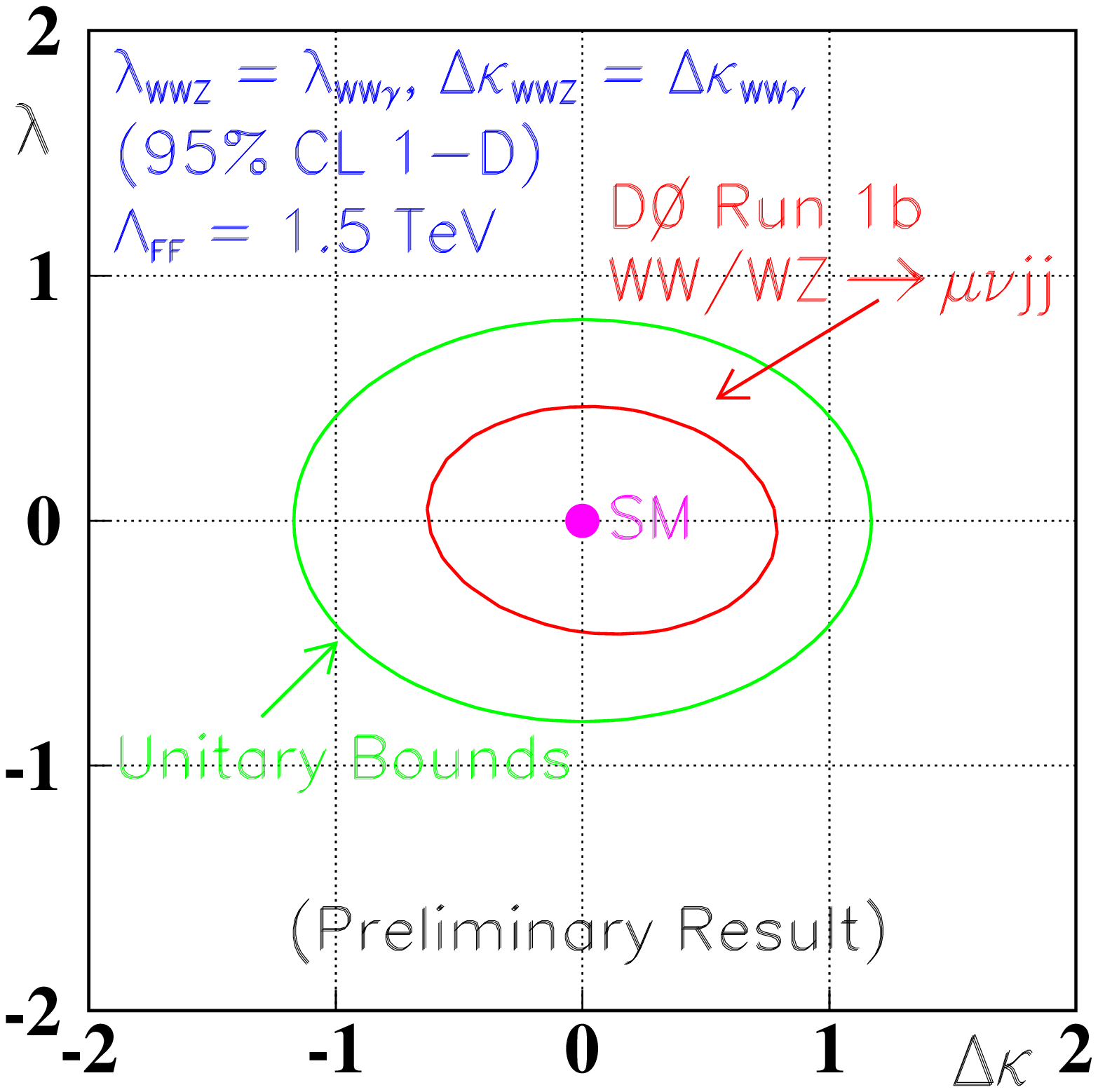}}
\vspace* {-0.2in}
\caption{Comparison of the final data sample and estimated total
background as function of $p_T(W)$ for $WW/WZ \to \mu\nu {\rm jj}$ is
shown on the left. On the right, the contour plot of allowed region in
the $\lambda-\Delta\kappa$ plane at 95\% C.L. for $\Lambda =$ 1.5
TeV. The outer ellipse shows the bounds imposed by the relevant
unitary conditions.}
\label{fig:mujet}
\end{figure}

The comparison of the final data sample and the total background as 
a function of $p_T(W)$ is shown in Figure~\ref{fig:mujet}. The data
and background
distributions are consistent with each other signalling no evidence of
AGC. This agreement is translated into AGC limits by means of a binned
maximum likelihood method with convoluting Gaussian errors for the
prediction and background uncertainities. The 95\% C. L. limits 
for the coupling parameters $\lambda$ and $\Delta\kappa$ are tabulated 
in Table~\ref{tab:mujet} for the cases of equal $WWZ$ and $WW\gamma$ 
couplings and the HISZ relations. The contour constraining the
coupling parameters in the $\lambda-\Delta\kappa$ plane is illustrated
in Figure~\ref{fig:mujet} for $\lambda_{WW\gamma}=\lambda_{WWZ}$  
and $\Delta\kappa_{WW\gamma}=\Delta\kappa_{WWZ}$.

\begin{figure}
\vspace*{-0.6in}
\hspace*{0.25in}
\epsfxsize 2.5in
\epsfysize 2.1in
{\epsfbox{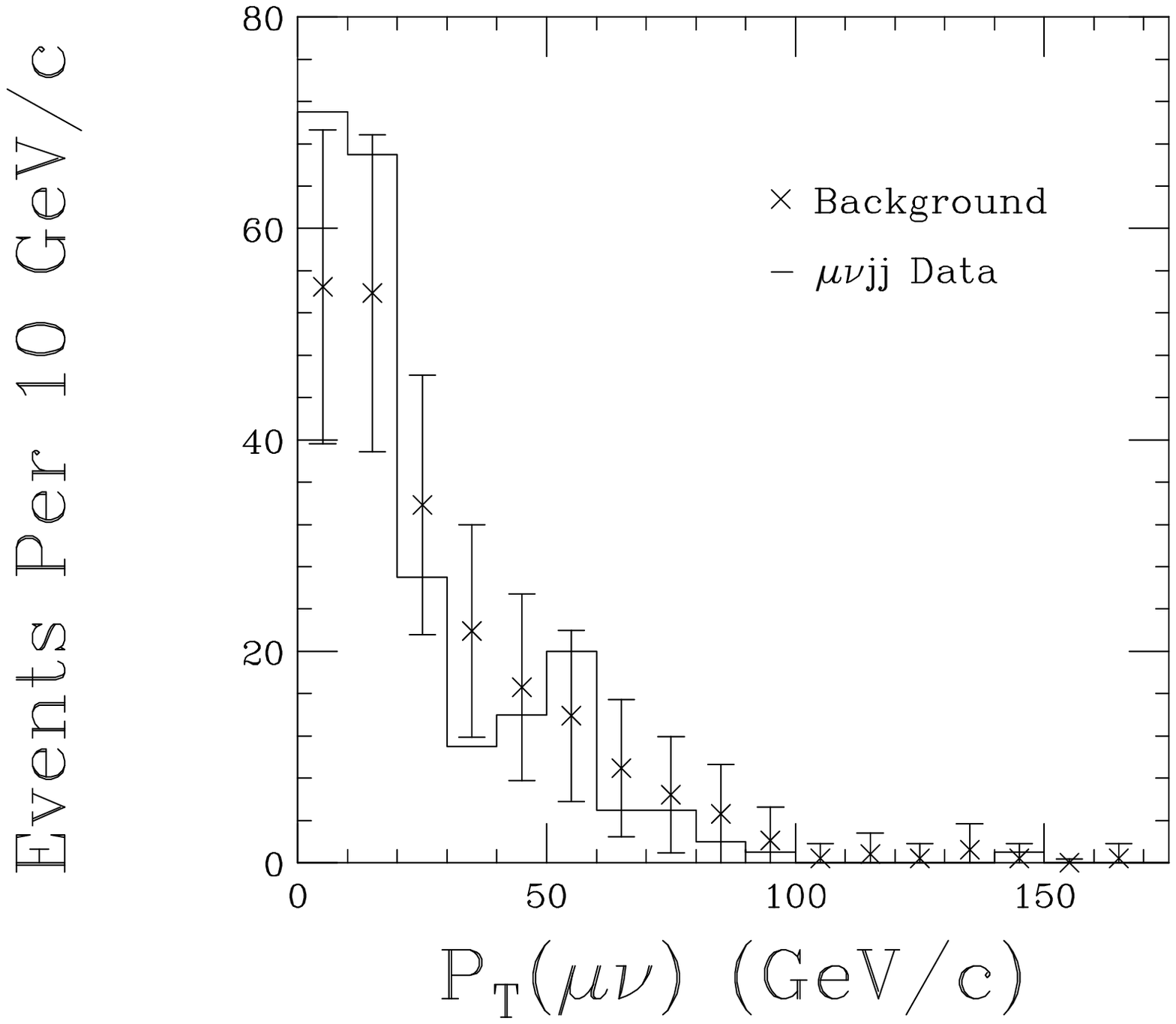}}
\end{figure}
\vspace* {-2.05in}

\subsection{ Overview of the \D0 Diboson Projects}

The \D0 collaboration has done a comprehensive search~\cite{ref:comp}
for anomalous
gauge couplings in the four diboson final states $p{\overline p} \to WW,
WZ, W\gamma, Z\gamma$ in twelve different decay modes using Run I
data. A detailed listing of the diboson final states and the relevant
decay channels including the status of the searches is given in 
Table~\ref{tab:overv}. The limits from all different channels have been 
combined to produce the final AGC coupling limits from \D0 in Run I.

\subsection{ Run I \D0 Combined Anomalous Coupling Limits}

The combined limit of $WW/WZ \to \mu\nu {\rm jj}$ channel along with
11 other channels listed in Table~\ref{tab:overv} has been produced. 
This involves performing
a simultaneous binned maximum likelihood fit to the observed number of
events and the expected number of signal and background distributions:
$p_T(\gamma)$ spectrum in the $W\gamma$ channels,
$p_T(l)$ distribution in $WW \to l\nu l\nu$ modes, $p_T(W \to l\nu)$ 
spectrum in the $WW/WZ \to l\nu {\rm jj}$ channels and to the observed 
number of events in the $WZ \to l\nu ee$ after a careful account of
correlated and uncorrelated uncertainities in different data sets and
modes. The one-dimensional 95\% C.L. axis limits for the various
coupling cases are given in Table~\ref{tab:comb}.

\begin{table}
\caption{A summary of the comprehensive search for anomalous
trilinear gauge couplings by \D0 using Run I data in 12 different 
channels.}
\label{tab:overv}
\begin{tabular}{llll}
{Diboson State} & Channel & {Coupling} & Status \\ \tableline \tableline
$Z\gamma$ & $\to ee\gamma, \mu\mu\gamma$ & $Z\gamma\gamma, ZZ\gamma$ &
Published \\
          & $\to \nu\nu\gamma$ &                &'93-94 Data:
published ('94-95 Data: in progress) \\ \tableline
$W\gamma$ & $\to e\nu\gamma, \mu\nu\gamma$   & $WW\gamma$  & Published \\
$WW$ & $\to e\nu e\nu, e\nu\mu\nu, \mu\nu\mu\nu$ & $WW\gamma, WWZ$ & 
Published \\ \hline
$WZ$ & {$\to e\nu ee, \mu\nu ee$} & {$WWZ$} & {Completed} \\ \tableline
$WW/WZ$ & $\to e\nu {\rm jj}$   & $WW\gamma, WWZ$  & Published \\
        & {$\to \mu\nu {\rm jj}$} &  & {Completed} \\ \tableline
Combined Limit & All completed & $Z\gamma\gamma, ZZ\gamma, WW\gamma,
WWZ$ & {Completed except '94-95 $Z(\nu\nu)\gamma$} \\
\end{tabular}
\end{table}
\section{Summary and Outlook}

The CDF and \D0 collaborations have done a number of world class 
measurements of electroweak
parameters using 1992-96 data such as $R_{e+\mu} = 10.48 \pm 0.43$, 
$\Gamma_W = 2.126 \pm 0.092$ GeV (indirect), $M_W = 80.43 \pm 0.11$
GeV, $g^W_\tau/g^W_e = 1.004 \pm 0.032$. The diboson channels $WW/WZ
\to \mu\nu {\rm jj}$, $WZ \to \mu\nu ee,\ e\nu ee$ have been
completed. Combined with other \D0 channels previously announced,
these provide some of the most stringent limits on the anomalous 
coupling parameters. Looking
towards future, the Main Injector is close to operation. The CDF and \D0
detectors are undergoing major upgrades. We expect much larger data
sets with much improved detectors in Run II. Both collaborations are
looking forward to an exciting Run II with many precision measurements
and new discoveries.

\begin{table}[t]
\caption{The \D0 Run I combined 1-D limits at 95\%
C.L. from a simultaneous fit to the $W\gamma,\ WW \to$ dileptons,
$WW/WZ \to e\nu{\rm jj}, \mu\nu {\rm jj}$, and $WZ \to$ trilepton data
samples for four different relations between $WWZ$ and
$WW\gamma$ couplings.}
\label{tab:comb}
\begin{tabular}{lcc}
{95{\%} C.L. Limits} & $\Lambda=1.5$ TeV & {$\Lambda = 2.0$ TeV} \\ 
\tableline \tableline
$\lambda_\gamma=\lambda_Z{\ }(\Delta\kappa_\gamma=\Delta\kappa_Z
= 0$) & { -0.20, 0.20}  & {-0.18, 0.19} \\ 
{$\Delta\kappa_\gamma=\Delta\kappa_Z{\ }(\lambda_\gamma=\lambda_Z
= 0$)} & { -0.27, 0.42}  & {-0.25, 0.39} \\ \tableline
{$\lambda_\gamma({\rm HISZ}){\ }(\Delta\kappa_\gamma
= 0$)} & { -0.20, 0.20}  & {-0.18, 0.19} \\ 
{$\Delta\kappa_\gamma({\rm HISZ}){\ }(\lambda_\gamma
= 0$)} & {-0.31, 0.56}  & {-0.29, 0.53} \\ \tableline
{$\lambda_Z({\rm SM}\ WW\gamma){\ }(\Delta\kappa_Z=\Delta g^Z_1
= 0$)} & {-0.26, 0.29} & {-0.25, 0.27} \\ 
{$\Delta\kappa_Z({\rm SM}\ WW\gamma){\ }(\lambda_Z=\Delta g^Z_1
= 0$)} & {-0.37, 0.55}  & {-0.34, 0.51} \\ 
{$\Delta g^Z_1({\rm SM}\ WW\gamma){\ }(\lambda_Z=\Delta\kappa_Z
= 0$)} & { -0.46, 0.65}  & {-0.44, 0.61} \\ \tableline
{$\lambda_\gamma({\rm SM}\ WWZ){\ }(\Delta\kappa_\gamma
= 0$)} & { -0.27, 0.25}  & {-0.25, 0.24} \\ 
{$\Delta\kappa_\gamma({\rm SM}\ WWZ){\ }(\lambda_\gamma
= 0$)} & { -0.57, 0.74}  & {-0.54, 0.69} \\
\end{tabular}
\end{table}

\section{Acknowledgement}

This work would not have been possible without the help of my \D0 and
CDF colleagues. This work is also supported in part by the U.S. 
Department of Energy Grant DE-FG03-94ER40837.


\begin{references}  
\vspace* {-0.5in}

\bibitem{ref:cdf}F. Abe {\sl et. al.} (CDF Collaboration),
Nuc. Inst. Meth. Phys. Res. {\bf A271}, 387 (1988).
\bibitem{ref:d0e}S. Abachi {\sl et. al.} (\D0 Collaboration),
Nuc. Inst. Meth. Phys. Res. {\bf A338}, 185 (1994).
\bibitem{ref:hamvan} R. Hamburg {\sl et al.} Nucl. Phys. {\bf B359},
343 (1991); W. L. van Neervan and E. B. Zijlstra, Nucl. Phys. {\bf
B382}, 11 (1992).
\bibitem{ref:seno} S. Eno for \D0 Collaboration, Talk presented at the
Les Renontres de Physique de la Vallee d' Aoste, LaThuille, Italy,
February 28 - March 6, 1999. The latest CDF results can be seen at 
{\sl http://www-cdf.fnal.gov/physics/ewk/ewk.html}.
\bibitem{ref:wnew} S. Abachi {\sl et. al.} (\D0 Collaboration), 
Phy. Rev. Lett. {\bf 75}, 1456 (1995).
\bibitem{ref:nutev} K. S. McFarland {et al.}, Proc. of the XXXIII
Rencontre de Moriond, March 1998, Les Arcs, France.
\bibitem{ref:resum} P. B. Arnold, R. P. Kaufman, Nucl. Phys. {\bf
B349}, 381 (1992); J. Collins, D. Soper, and G. Sterman,
Nucl. Phys. {\bf B250}, 199 (1985); R. K. Ellis and S. Veseli,
Nucl. Phys. {\bf B511}, 649 (1998); R. K. Ellis {\sl et al.}, 
Nucl. Phys. {\bf B503}, 309 (1997).  
\bibitem{ref:pvt} Robert Wagner, private communication.
\bibitem{ref:wptd0} M. Mostafa for \D0 Collaboration, Talk presented
at the APS Meeting, Atlanta, GA, March 21-26, 1999. See also at {\sl  
http://www-d0.fnal.gov/public/d0\_physics\_v2.html}.
\bibitem{ref:agcl} K. Hagiwara, R. D. Peccei, D. Zeppenfeld, and
K. Hikasa, Nucl. Phys. {\bf B282}, 2253 (1987); K. Hagiwara,
J. Woodside, and D. Zeppenfeld, Phys. Rev. {\bf D41}, 2113 (1990).
\bibitem{ref:hisz} K. Hagiwara {\sl et al.}, Phys. Rev. {\bf D48},
2182 (1993).  
\bibitem{ref:kggg} K. Gaemers and G. Gounaris, Z. Phys. {\bf C1}, 259 
(1979). 
\bibitem{ref:fake} B. Abbot {\sl et al.}, (\D0 Collaboration),
Phys. Rev. {\bf D54}, 052001 (1998).
\bibitem{ref:herw} G. Marchesini {\sl et al.}, Phys. Comm. {\bf 67},
465 (1992). 
\bibitem{ref:gean} F. Carminati {\sl et al.}, GEANT Users Guide, CERN
Program Library Long Writeup WSO13 91993), unpublished.
\bibitem{ref:comp} S. Abachi {\sl et al.}, (\D0 Collaboration), 
Phys. Rev. Lett., {\bf 75}, 1023 (1995); Phys. Rev. Lett., {\bf 75}, 
1028 (1995); Phys. Rev. Lett., 
{\bf 75} 1034(1995); Phys. Rev. Lett., {\bf 77}, 3303 (1996);
Phys. Rev. Lett., {\bf 78}, 3634 (1997); Phys. Rev. Lett., {\bf 781},
3640 (1997); Phys. Rev., {\bf D56}, 6742 (1997); S. Abbott {\sl et al.},  
(\D0 Collaboration), Phys. Rev. Lett., {\bf 79}, 1441 (1997);
Phys. Rev., {\bf D57} 3817 (1998); Phys. Rev., {\bf D58}, Rapid
Comm. 051101 (1998); Phys. Rev., {\bf D58}, Rapid
Comm. 31102 (1998).

\end{references}
\end{document}